\documentclass[12pt]{article}
\usepackage[utf8]{inputenc}
\usepackage{amsmath, amssymb}
\usepackage{graphicx}
\usepackage{booktabs}
\usepackage{placeins}
\usepackage{subcaption}
\usepackage{bbm}
\usepackage{pgfplots}
\pgfplotsset{compat=1.17}
\usepackage{hyperref}

\title{Scaling Laws for Economic Productivity: Experimental Evidence in LLM-Assisted Translation\thanks{I gratefully acknowledge financial support from Open Philanthropy to conduct this research. The research described in this article was approved by the Yale Human Research Protection Program and was preregistered at the AEA RCT Registry (AEARCTR-0013743). We thank Charlie Harrison for his invaluable research assistance on this experiment and Jason Abaluck and Pascual Restrepo for comments and suggestions.}}
\author{
    Ali Merali\textsuperscript{\textdagger} \\ 
}

\begin{document}
\maketitle

\renewcommand{\thefootnote}{\textdagger} 
\footnotetext{Yale University, Department of Economics: ali.merali@yale.edu}

\renewcommand{\thefootnote}{\arabic{footnote}} 

\begin{abstract}
This paper derives "scaling laws"—empirical relationships between the training compute of Large Language Models (LLMs) and their performance—for economic outcomes. In a preregistered online experiment, 300 professional translators completed 1,800 tasks using one of 13 LLMs (or a control). A tenfold increase in model compute improved task completion speed by 12.3\%, grades by 0.18 standard deviations, and earnings per minute by 16.1\%. Gains were four times larger for lower-skilled workers. These findings suggest continued model scaling could boost U.S. productivity by at least 6.9\% over the next decade.

\end{abstract}

\pagebreak

\section{Introduction} 

The amount of training compute used by frontier large language models (LLMs) increased by roughly 5000x between the release of GPT-2 in 2019 and GPT-4 in 2023. Estimates from Epoch AI, an AI research institute, suggest a similar increase over the next six years. How does this massive increase in model training compute, however, map onto performance? The empirical machine learning literature has derived remarkably consistent ‘scaling laws’ suggesting a strong relationship between a model’s training compute and model perplexity, a measure of model loss, across more than seven orders of magnitude. But there is so far a very limited understanding of how this reduction in perplexity affects key economic and social outcomes.  \\\

This paper aims to offer the first experimental evidence on this question. It conducts a randomized controlled trial (RCT) involving 300 professional translators conducting 1800 tasks of varying complexities. The participants were randomly assigned to either treatment groups where they could utilize one of thirteen LLMs of differing model training compute to help them complete their task or to a control group where they completed tasks without any AI assistance. Participants faced high-powered incentives with significant bonus payments for high-quality tasks as evaluated by three experienced professionals in the field. The key outcome variables, therefore, were how translator’s time taken, quality of tasks completed, and earnings per minute (inclusive of bonuses) varied by model training compute. \\\

The results were stark. For every 10x increase in model training compute, translator’s managed to complete a task 12.3\% quicker (p=0.001). Converted into ‘GPT-jumps’, denoting the ~70x difference in model training compute between successively numbered GPT models, this equates to a 22.7\% decrease in time per model jump. Not only were responses completed more quickly, however, but they also received higher grades. For every 10x increase in model compute grades improved by 0.25 points on a 7-point scale or by 0.18 standard deviations (p=0.000). This means that the translator’s overall productivity improvements (measured in earnings per minute, inclusive of bonuses) was even greater than the reduction in time taken alone with a 16.1\% increase in earnings per 10x increase in model compute or a 29.7\% increase per ‘GPT-jump’ (p=0.001). \\\

These gains were not split evenly between all translators. All participants completed a baseline task without any AI assistance offering a metric for translator skill. Splitting the translator’s into two groups based on whether their first task was completed in above or below the median time the impact of increased model compute differed significantly. Those with ‘high-skill’ reduced their time taken per task by 4.9\% whilst those with ‘low-skill’ saw a more than 4x larger reduction of 21.1\% (p=0.017). These results have potentially significant implications for the degree to which future AI improvements are skill-biased and their subsequent impacts on wage inequalities. \\\

Utilizing the experimental findings on economic scaling laws, I estimate the aggregate productivity gains from AI on the US economy over the next decade. This calculation mirrors that of Acemoglu (2024) and utilizes the same framework based on Hulten's theorem- the only changes are to allow for the possibility of model improvement as the scaling hypothesis suggests, and to take into account recent data on inference costs. These changes leads to a significantly higher productivity estimate of around a magnitude higher than in Acemoglu (2024) with an estimate 6.9\% productivity growth. \\\

The rest of the paper is organized as follows. Below, connections to four existing academic literatures are drawn. In Section Two, an overview of the experimental design is offered alongside some summary statistics on how the translator’s viewed the experimental design. This includes the degree to which they considered the tasks as similar to ones they experienced as professionals and their previous familiarity with AI tools. In Section Three, all the results are presented. In Section Four, a basic quantitative estimate is made for the aggregate productivity gains from AI over the next decade. Finally, in Section Five the paper concludes with discussion on the results. \\\

There’s already a wide range of existing economic evidence suggesting that current frontier LLMs are able to lead to significant productivity enhancements across varied tasks. For instance, Noy and Zhang (2023) found that ChatGPT can lead to productivity improvements of 37\% for a range of professional writing tasks including data analysis and marketing. These are similar estimates to the double-digit productivity improvements found in code completions with GitHub Copilot (Kalliamvakou et al 2022) and legal tasks (Choi et al 2023) with GPT-4. Research using BCG consultants to perform professional tasks, however, highlights the potential for significant variation with GPT-4 yielding double-digit productivity enhancements on some common consulting tasks whilst offering no such benefits on others. These papers, alongside many others, show the impact of workers being offered an AI model or not. This paper aims to extend this literature by plotting out the trajectory of economic impacts as AI model sizes grow and automation capabilities increase. \\\

Secondly, the relationship between how model capabilities increase as the amount of compute used to train the model increases is best understood through ‘scaling laws’ such as those most prominently derived by Kaplan et al (2020). These empirical scaling laws found a remarkably consistent trend in the change in cross-entropy loss over seven orders of magnitude of compute. There exists a significant machine learning literature now aimed at understanding this relationship both in LLMs and other AI domains such as vision, RL, and speech. In the social science literature, an experiment by Hackenburg et al (2024) finds that similar such scaling laws can be derived for purely machine performance on persuasion tasks. There is, however, a more limited understanding currently of how such changes in the cross-entropy loss of an LLM affect human performance on tasks and how these lead to changes in important economic variables such as productivity or skill-based wage premia. \\\

Thirdly, this paper also aims to contribute to a literature in labor economics aiming to understand the relationship between technological growth and skill-biased wage premia. Whether a given technology is skill-biased or not can have significant implications on changes in wage inequality as depicted by research by Goldin (2007). Preliminary experimental evidence from research in professional tasks as well as in the legal field suggests that using an LLM offers the greatest productivity enhancement to those who are lowest-skilled. This study aims to extend this literature by considering a continuous measure of automation to better understand how the skill-based wage premia changes according to increases in LLM capabilities. \\\

Finally, this paper adds to the literature on the effect of machine translation on a translator's productivity and quality. This literature has existed since 2013 (Kalchbrenner and Blunsom, 2013) and has seen a variety of machine tools used based on varying different architectures from neural machine translation (NMT) to the more recent transformer architecture. A good early literature review is provided by Sanchez and Torron (2017). This paper contributes to this literature by analyzing the effects of 13 different LLMs of different compute sizes on translator productivity and quality. \\\

\section{Experimental Design}

For this experiment, three hundred participants were recruited through online recruiting platforms Freelancer and Fiverr. These participants were split evenly based on their proficiency on the three languages of interest for this study Spanish, Hindi, and Arabic (all translations were made from English into the target language). All the participants completed a pre-screening survey in order to be eligible to take part in the full survey; requirements included that they had at least one year of experience as a professional translator, that they completed paid translation tasks in the past year, that they met certain standards of language ability, and that they were comfortable having their AI model usage monitored throughout the experiment. \\\

Over 70\% of eligible participants had at least three years of professional experience as a translator and just over a third had at least five years of experience as a professional. In self-reported analysis, translators judged their translation abilities between English and the target language at 4.52/5 and their professional fluency in the desired language at 4.85/5. Further, they reported a familiarity of AI tools for translation of 4.15/5 and self-evaluated their abilities in using AI tools for translation at 4.18/5. \\\

Participants who met all eligibility criteria then completed a baseline task where they translated a text from English without any AI or software assistance. This task served as a benchmark of translation ability that was used as a metric for the translator’s skill. If the translation wasn’t completed to a sufficiently high quality the participant’s results for later tasks were not included in the experiment results. \\\

The participants then completed five more tasks during the experiment. The tasks were all short, taking almost exactly ten minutes to complete each on average. The texts to be translated covered a range of materials including business, academic, and literary texts to simulate a wide variety of potential professional use cases. Participants overall reported that the tasks had an overall similarity score with those they completed professionally of 3.53 out of 5. A full description of the tasks involved can be found in Appendix B. \\\

Participants were given high-powered incentives to complete high-quality tasks. In return for completing the full survey all participants were paid \$10. They could earn up to \$12 more, however, in bonus payments depending on performance. These were paid out as \$2 per task that they achieved a score of six points or higher out of seven. These grades consisted of the average of three professional translators who went through more stringent vetting (they had an average of 5+ years of experience each, and were monitored in their grading of sample tasks). The graders themselves were incentivized to grade carefully by offering significant bonus payments (up to doubling their payment) based on if their grades given were sufficiently close to the average grade given by other graders. \\\

To help the participants with their tasks, they were assigned either one of thirteen AI models to aid them (treatment) or were told to complete the tasks by themselves (control). The participants were then given access to the desired model and their interactions could be monitored. Before each genuine task they were given a practice task to use the model in so that they were able to understand the performance of the AI model they had been assigned and so that compliance with the experiment could be monitored. \\\

\section{Results}
\textbf{Section 3.1: Productivity and Quality Impacts of any AI Usage} \\\

As a preliminary analysis, we study the impacts of participants receiving any AI model versus the control. All thirteen AI models of differing compute sizes are therefore pooled into a single treatment group. As depicted by Figure 1 below, the average time taken by participants to conduct a task without access to any AI model was almost exactly ten minutes (600.7 seconds). With access to any AI model this was reduced to 413.8 seconds (p=0.000), a statistically significant reduction in time of 31.1\%. \\\

Not only were participants much quicker in completing tasks with access to AI models, however, but their tasks were also associated with higher quality scores although this result was not statistically significant (p=0.148)  as shown in Figure 1. The average grade without access to an AI model was 4.51. When participants were able to use an AI model this increased to 4.71, an increase of 0.2 points on a seven-point scale. This equates to an increase of 0.14 standard deviations in the average grade. Participants therefore completed tasks both significantly quicker and with higher quality when given access to any AI model. Both of these results are reported in regression tables with and without controls as Table 1 in Appendix A. \\\

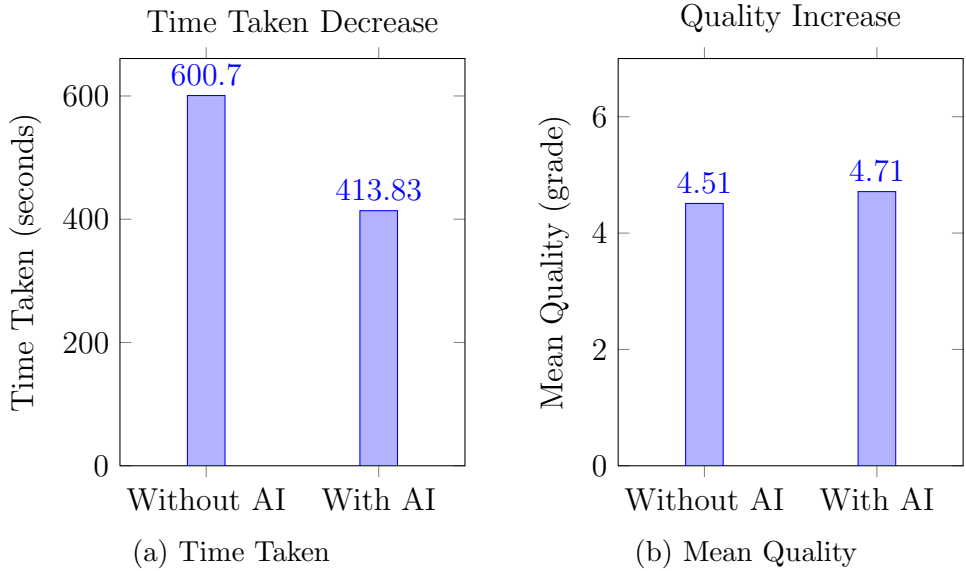
\begin{figure}[ht]
    \centering
    \begin{subfigure}[b]{0.45\textwidth}
        \centering
        \begin{tikzpicture}
        \begin{axis}[
            ybar,
            bar width=0.5cm,
            symbolic x coords={Without AI, With AI},
            xtick=data,
            nodes near coords,
            ymin=0,
            ylabel={Time Taken (seconds)},
            title={Time Taken Decrease},
            enlarge x limits=0.5,
            width=\textwidth, 
            height=7cm, 
            ymin=0 
            ]
            \addplot coordinates {(Without AI, 600.70) (With AI, 413.83)};
        \end{axis}
        \end{tikzpicture}
        \caption{Time Taken}
    \end{subfigure}
    \hspace{0.4cm} 
    \begin{subfigure}[b]{0.45\textwidth}
        \centering
        \begin{tikzpicture}
        \begin{axis}[
            ybar,
            bar width=0.5cm,
            symbolic x coords={Without AI, With AI},
            xtick=data,
            nodes near coords,
            ymin=0,
            ymax=7, 
            ylabel={Mean Quality (grade)},
            title={Quality Increase},
            enlarge x limits=0.5,
            width=\textwidth, 
            height=7cm 
            ]
            \addplot coordinates {(Without AI, 4.509) (With AI, 4.713)};
        \end{axis}
        \end{tikzpicture}
        \caption{Mean Quality}
    \end{subfigure}

    \caption{Impact on Time Taken and Quality of any AI Model Usage}
\end{figure}

\FloatBarrier 

\textbf{Section 3.2: Scaling Laws for Time Taken} \\\

We now move on to deriving the impacts of additional model training compute on the productivity of participants. These results are summarized in Figure 2. When participants were given access to a model with an increase in the amount of training compute of 10x this was associated with a 12.3\% reduction in time in completing tasks (p=0.001). This implies that a ‘GPT-jump’ in models, denoting the roughly 70x increase in training compute between successively numbered GPT model versions, is associated with a 22.7\% decrease in the amount of time spent by participants. The full regression table corresponding to this result can be found as Table 2 in Appendix A. \\\

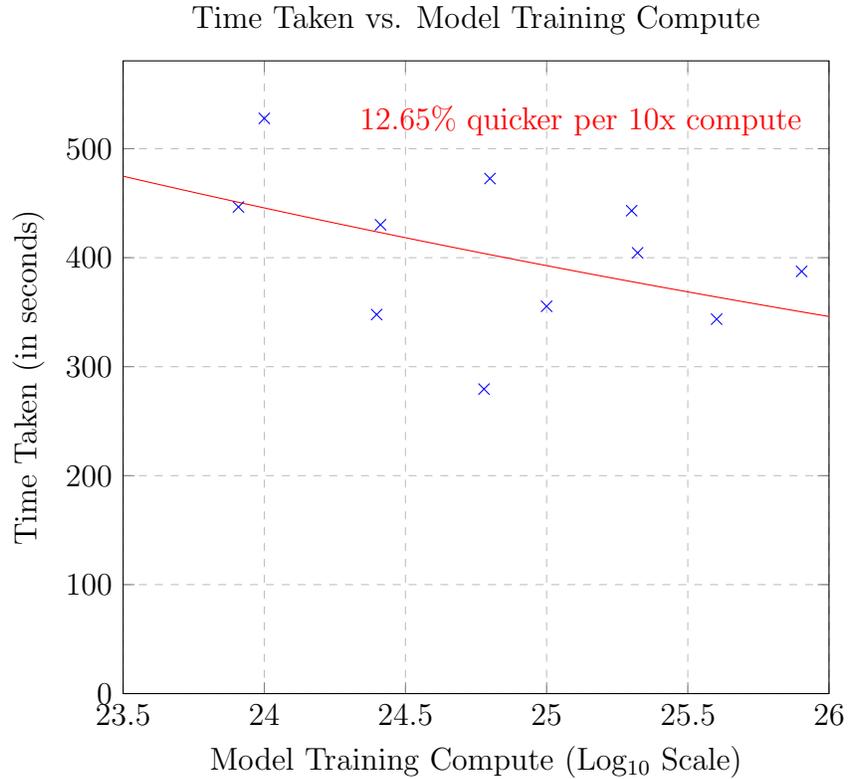
\begin{figure}[ht]
    \centering
    \begin{tikzpicture}
    \begin{axis}[
        xlabel={Model Training Compute (Log$_{10}$ Scale)},
        ylabel={Time Taken (in seconds)},
        xmin=23.5, xmax=26, 
        ymin=0, 
        ymajorgrids=true,
        xmajorgrids=true,
        grid style=dashed,
        height=10cm, 
        width=0.8\textwidth, 
        title={Time Taken vs. Model Training Compute},
        mark options={scale=1.5,blue}, 
        ]
        \addplot[
            only marks,
            mark=x,
            mark options={scale=1.5,blue}, 
            ] coordinates {
            (22.924280166625, 543.1563)
            (23.908485412597, 446.5391)
            (24, 527.8431)
            (24.397939682006, 347.8256)
            (24.411619186401, 430.2198)
            (24.778150558471, 279.4401)
            (24.799341201782, 472.6224)
            (25, 355.4247)
            (25.301029205322, 443.134)
            (25.322219848632, 404.4511)
            (25.602060317993, 343.603)
            (25.903089523315, 387.4423)
        };
        
        \addplot [
            domain=23.5:26, 
            samples=100, 
            color=red
        ] {exp(9.1356 - 0.1265 * x)}; 
        
        \node at (axis cs:24.3,550) [anchor=north west,red] {12.65\% quicker per 10x compute};
        
    \end{axis}
    \end{tikzpicture}
    \caption{Time Taken as a Function of Model Training Compute (Log Scale)}
\end{figure}

\FloatBarrier 

\textbf{Section 3.3: Scaling Laws for Quality} \\\

The above result showed that giving participants access to models with larger training computes led them to complete tasks significantly quicker. Is this reduction in time spent associated with a decrease in task submission quality? No. Instead, as summarized in Figure 3 (and in Table 3 of Appendix A), participants scored on average 0.25 points higher (on a 7 point grading scheme) when given a model with training compute size an order of magnitude greater (p=0.000). This is equivalent to a 0.18 standard deviation improvement in grades. A ‘GPT-jump’ of a 70x increase in model compute is therefore associated with a 0.33 standard deviation increase in grades. \\\

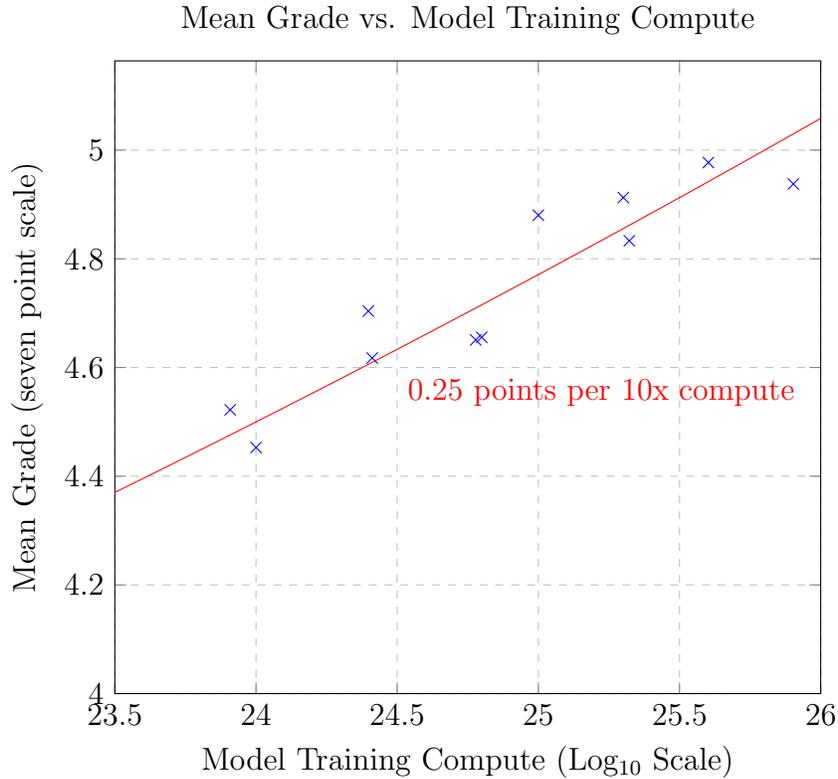
\begin{figure}[ht]
    \centering
    \begin{tikzpicture}
    \begin{axis}[
        xlabel={Model Training Compute (Log$_{10}$ Scale)},
        ylabel={Mean Grade (seven point scale)},
        xmin=23.5, xmax=26, 
        ymin=4, 
        ymajorgrids=true,
        xmajorgrids=true,
        grid style=dashed,
        height=10cm, 
        width=0.8\textwidth, 
        title={Mean Grade vs. Model Training Compute},
        mark options={scale=1.5,blue}, 
        ]
        \addplot[
            only marks,
            mark=x,
            mark options={scale=1.5,blue}, 
            ] coordinates {
            (22.924280166625, 4.188257)
            (23.908485412597, 4.522121)
            (24, 4.452802)
            (24.397939682006, 4.703858)
            (24.411619186401, 4.617492)
            (24.778150558471, 4.650816)
            (24.799341201782, 4.655667)
            (25, 4.880216)
            (25.301029205322, 4.912686)
            (25.322219848632, 4.833165)
            (25.602060317993, 4.977161)
            (25.903089523315, 4.937725)
        };
        
        \addplot [
            domain=23.5:26, 
            samples=100, 
            color=red
        ] {exp(0.1001 + 0.0585 * x)}; 
        
        \node at (axis cs:24.5,4.6) [anchor=north west,red] {0.25 points per 10x compute};
        
    \end{axis}
    \end{tikzpicture}
    \caption{Mean Grade as a Function of Model Training Compute (Log Scale)}
\end{figure}

\FloatBarrier 

\textbf{Section 3.4: Scaling Laws for Productivity} \\\

The above results set the stage for the preferred pre-registered metric for task productivity, earnings per minute. Participants were offered a flat rate of \$10 for completing all six tasks to a satisfactory standard and were offered a \$2 bonus per task for a high-quality completion. A task was deemed of high-quality if it received an average grade (from the three human expert graders) of 6 or higher whilst a task was considered satisfactory if it received an average grade of 2 or higher. The average participant earned \$1.18 per minute of completing tasks. This increased by \$0.19 a minute for every 10x increase in compute the model they received was given in training (p=0.001). As such, the average participant gained an 16.1\% increase in earnings for every 10x increase in model training compute or a 29.7\% increase for every ‘GPT-jump’. These results are summarized in Figure 3 below and a full regression output can be found as Table 4 in Appendix A. \\\

\begin{figure}[ht]
    \centering
    \begin{tikzpicture}
    \begin{axis}[
        xlabel={Model Training Compute (Log$_{10}$ Scale)},
        ylabel={Productivity (Earnings per Minute with Bonus)},
        xmin=23.5, xmax=26, 
        ymin=0.7, 
        ymajorgrids=true,
        xmajorgrids=true,
        grid style=dashed,
        height=10cm, 
        width=0.8\textwidth, 
        title={Productivity vs. Model Training Compute},
        mark options={scale=1.5,blue}, 
        ]
        \addplot[
            only marks,
            mark=x,
            mark options={scale=1.5,blue}, 
            ] coordinates {
            (22.924280166625, 0.9141558)
            (23.908485412597, 1.08585)
            (24, 0.7959957)
            (24.397939682006, 1.362565)
            (24.411619186401, 0.9356401)
            (24.778150558471, 1.372728)
            (24.799341201782, 0.9271468)
            (25, 1.296351)
            (25.301029205322, 1.385925)
            (25.322219848632, 1.171472)
            (25.602060317993, 1.676219)
            (25.903089523315, 1.310305)
        };
        
        \addplot [
            domain=23.5:26, 
            samples=100, 
            color=red
        ] {exp(-4.3213 + 0.1810 * x)}; 
        
        \node at (axis cs:24,1.5) [anchor=north west,red] {16.1\% epm increase per 10x Compute};
        
    \end{axis}
    \end{tikzpicture}
    \caption{Productivity as a Function of Model Training Compute (Log Scale)}
\end{figure}
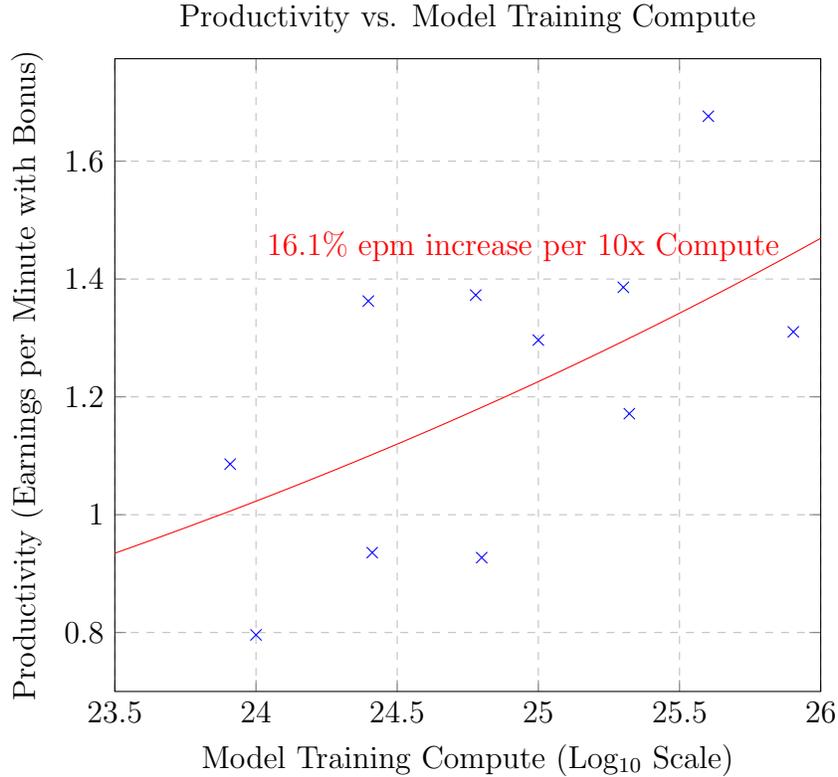

\FloatBarrier 

\textbf{Section 3.5: Scaling Laws for Skill-Based Wage Inequality} \\\

All participants completed a baseline task without access to any AI models for which they received the same fixed and bonus payments. As such, it is possible to rank participants on their earnings per minute (with bonuses) on this baseline task. In this section we analyze whether the participants who gain the most from AI models with greater training compute are those with the lowest or highest earnings on the baseline task- namely are LLMs skill-biased or not? \\\

The participants were split into two groups based on their performance on the baseline task. Those who completed the baseline task in below median time were classed as high-skill whilst though who took above median time were classed as low-skill. As the regression results in Figure 4 depict, there is a significant difference in the gains of using AI models with greater training compute by skill level. In particular, whilst high-skilled participants reduce their time taken by 4.9\% for every 10x increase in compute those with low-skill have a 21.1\% reduction (p=0.017). These translate into 9\%  and 39.0\% reductions in time taken per 'GPT-Jump'. Scaling LLMs therefore leads to gains with significant heterogeneity by participant skill, with lower-skilled participants obtaining the largest gains.  \\\

\begin{table}[htbp]\centering
\caption{Heterogeneity of Scaling Laws on Time Taken by Baseline SKill Level}
\resizebox{\textwidth}{!}{%
\begin{tabular}{l*{4}{c}}
\hline\hline
 & \multicolumn{2}{c}{(1)} & \multicolumn{2}{c}{(2)} \\
VARIABLES & \multicolumn{2}{c}{time} & \multicolumn{2}{c}{time} \\ \hline
 & Coefficient & Std. Err. & Coefficient & Std. Err. \\ \hline
Skill & 2,084.416*** & 722.866 & 2,015.373*** & 728.927 \\
 & (2.88) &  & (2.76) &  \\
Skill\_logmodelcompute & -69.399** & 29.141 & -66.523** & 29.386 \\
 & (-2.38) &  & (-2.26) &  \\
logmodelcompute & -20.845 & 20.449 & -22.725 & 20.642 \\
 & (-1.02) &  & (-1.10) &  \\
hindi & -22.599 & 30.038 &  &  \\
 & (-0.75) &  &  &  \\
arabic & 57.941 & 29.608 &  &  \\
 & (1.96) &  &  &  \\
task2 & 150.326*** & 38.930 &  &  \\
 & (3.86) &  &  &  \\
task3 & 79.537** & 38.747 &  &  \\
 & (2.05) &  &  &  \\
task4 & 172.353*** & 38.808 &  &  \\
 & (4.44) &  &  &  \\
task5 & 44.856 & 38.683 &  &  \\
 & (1.16) &  &  &  \\
\_cons & 649.382 & 507.939 & 796.711 & 511.689 \\
 & (1.28) &  & (1.56) &  \\
\hline
Observations & \multicolumn{2}{c}{1,392} & \multicolumn{2}{c}{1,392} \\
R-squared & \multicolumn{2}{c}{0.1671} & \multicolumn{2}{c}{0.1460} \\
Adj R-squared & \multicolumn{2}{c}{0.1617} & \multicolumn{2}{c}{0.1441} \\
\hline\hline
\multicolumn{5}{l}{\footnotesize Notes: *** p<0.01, ** p<0.05, * p<0.1}
\end{tabular}%
}
\end{table}

\section{Aggregate Productivity Gains}

In this section, we utilize the above results on the productivity gains from model scaling to estimate aggregate productivity gains from AI over the next ten years. In doing so, we leverage the framework from Acemoglu (2024). Here a version of Hulten's theorem is used to estimate aggregate productivity gains based on the fractions of tasks in the economy that are affected by AI and the average task-level cost savings. \\\

The estimates from Acemoglu (2024) therefore are based on the multiplication of four parameters. Firstly, an estimate of the share of tasks exposed to AI is used from Eloundou et al (2023) which estimates it at 19.9\%. Secondly, as some tasks involve a combination of labor and capital the (AI-exposure adjusted) labor share of 0.57 is used. Neither of these parameters are changed in the analysis of this section. \\\

However, the third parameter estimating the labor cost savings from AI is changed significantly. The estimate from Acemoglu (2024) come from two studies; Brynjolfsson et al (2023) which looked at a real-world case study involving call center operatives and Noy and Zhang (2024) who used an online experiment for a range of white-collar tasks. The estimate comes from an average of the two productivity effect sizes which is 27\%. Note that these estimates do not take into account the possibility that the productivity effects may grow over time as models scale in the amount of training compute used and model quality improves. Further, they combine estimates from two different generations of models; the lower estimate of 14\% from Brynjolfsson et al (2023) comes from the usage of a GPT-based model rolled out in 2020-2021 whereas the 40\% estimate in Noy and Zhang (2023) comes from the larger and higher quality ChatGPT model (both models have already been superseded by much larger and higher-performing models by OpenAI, Anthropic, Meta, Alphabet, and others in 2024 alone). \\\ 

In particular, the 14\% estimate comes from a model estimated to be trained on the order of $10^{23}$ training FLOPs of model compute whilst the 40\% estimate comes from a model estimated to be trained on the order of $10^{24}$ training FLOPs. This compares to models trained on the order of $10^{25}$ training FLOPs in this study. It is likely that if either of these studies was completed on the best available models today the productivity effects would be significantly higher - how can we estimate just how large we expect these productivity estimates to be? \\\ 

Here we use evidence from the scaling results from this experiment. Note we do not assume that the scaling laws for different tasks all follow those of the scaling laws for translation- other tasks may see differential productivity gains. The weaker assumption that is made here, however, is that the ratio of the productivity gains made from going from no AI assistance to assistance from an LLM with training compute x to the productivity gains from going from assistance with an LLM with training compute x to an LLM with a larger training compute y for any given task is the same as it is for translation. \\\

Note that Eloundou et al (2023) estimate translators to be the occupation with either the highest or close to the highest (depending on the model they use) AI-exposed occupation. As such, we would expect the ratio of gains from AI-assistance to be disproportionately larger from smaller models to larger models for translators than for other tasks. This assumption, therefore, likely yields an underestimate for the scaling effects of all other tasks. \\\

Nonetheless, evidence from Table 2 suggests that for translation 54.9\% of the gains from using the largest AI model in this study, Claude-3.5 Sonnet, would be received by a model of the size used in Brynjolfsson et al (2023) and 72.8\% of the gains from the model of the size used in Noy and Zhang (2023). This therefore estimates that if these studies were replicated with the largest currently available models their productivity effects would be 25.5\% and 54.9\% respectively for an average of 40.2\%. \\\

Of course, the current largest models are extremely unlikely to be the largest models available in ten years time. Conservatively, we use Epoch AI's estimates of models with  the order of $10^{30}$ training FLOPs being commercially available by 2030 and allow the remaining time for their widespread adoption. Again utilizing evidence from Table 2, and assuming the ratio of progress between the jump from no models to smaller models and the jump from smaller models to larger models mirrors the estimates for translation, we estimate that replications of Brynjolfsson et al (2023) with the best models available in 2030 would have a 39.1\% productivity boost whilst similar replications of Noy and Zhang (2023) would see a 84.4\% boost. This yields an average of a 61.2\% productivity effect in ten years time when we account for (both past and) future model scaling, compared to the original 27\% estimate. \\\

Finally, the remaining parameter in Acemoglu's estimate is whether tasks that can be automated can be done so profitably. Here, he uses an estimate from Svanberg et al (2024) suggesting 23\% of automatable tasks would be economically feasible to do so within then years. However, this estimate is for computer vision tasks which can involve high fixed costs such as camera equipment. However, the cost can be substantially smaller for tasks that can be automated through LLMs. Indeed the inference costs for all 300 translators completing nearly 1,400 tasks with AI assistance combined came to significantly less than a dollar. \footnote{This ignores the time cost of a professional interacting with the LLM interface, by far a more significant cost than inference costs, however this is already accounted for in the time-based productivity estimates} For this domain therefore, the profitable automation rate would be roughly 100\%. \\\

More generally, the inference cost of one of the leading models as of the time of writing is \$1 for $100,000$ syllables of writing. As such, if any task can be automated via an LLM it is almost certain it will be profitable to do so based on inference costs. Further, a report from Andreesen Horowitz, a venture capitalist firm, shows that the inference cost of an LLM achieving a certain fixed benchmark score has recently been falling by 10x per year. If it wouldn't currently be profitable to use an LLM to aid for a task, therefore, it is likely to be so in the near-term future even if such a rapid rate of cost decreases doesn't continue. This means that for the purpose of this calculation we assume that if LLMs can increase productivity on a task it will always be profitable. \\\

This means that the total productivity gains from AI over the next ten years is estimated through the multiplication of three parameters. The share of tasks that can be automated (19.9\%) multiplied by the average productivity effect using scaling laws (61.2\%) multiplied by the labor share of costs (57\%). This yields an estimate of 6.95\% productivity growth over the next decade from LLMs, a significant amount of growth. \\\

It should be noted that this calculation makes several restrictions. For instance, it is assumed that the task structure of the economy doesn't change at all despite significant technological changes. Further, and of particular significance, it also assumes no change to the rate of technological progress itself. There is already a wealth of evidence, however, that AI can increase the pace of scientific discovery itself for instance in predicting protein structures (Jumper et al 2021) or discovering new materials technology (Toner-Rodgers, 2024). This can lead to much higher estimates such as in Korinek and Suh (2024), therefore there is a sense in which these estimates may be considered a lower bound. \\\

\section{Discussion}

Previous research has already highlighted the potential for currently available LLMs to offer double-digit productivity improvements in a wide range of settings including legal tasks, consulting, software engineering, and a variety of professional tasks. Combined with forecasts from Epoch AI that frontier LLMs may use 10,000x more training compute in six years time than they currently do this experimental evidence suggests that future generations of LLMs may have significant productivity improvements. These improvements may be heterogenous with respect to ability, however, with the lowest-skilled workers reaping the greatest gains. \\\

There are many limitations to this study. This paper focused on a single professional skill (translation) and only tested participants on short tasks. Further, the results were only derived on a range of just over two orders of magnitude of compute. \\\

Whether these economic scaling laws generalize to other domains and for greater model training compute sizes is a question for further research. For now, the evidence provided suggests future frontier LLMs may have significant economic implications. \\\

\pagebreak

\section*{Bibliography}

\begin{itemize}

\item Acemoglu, Daron. "The Simple Macroeconomics of AI." National Bureau of Economic Research, Working Paper Series, Working Paper No. 32487, May 2024. .

\item Acemoglu, Daron, and Pascual Restrepo. "Robots and Jobs: Evidence from US Labor Markets." Journal of Political Economy, 2020, 128 (6), 2188–2244.

\item Acemoglu, Daron, and Pascual Restrepo. "The Race between Man and Machine: Implications of Technology for Growth, Factor Shares, and Employment." American Economic Review, 2018, 108 (6), 1488–1542.

\item Agarwal, Nikhil, et al. Combining human expertise with artificial intelligence: Experimental evidence from radiology. No. w31422. National Bureau of Economic Research, 2023.

\item Brynjolfsson, Erik, Danielle Li, and Lindsey R. Raymond. Generative AI at work. No. w31161. National Bureau of Economic Research, 2023.

\item Choi, Jonathan H., Amy Monahan, and Daniel Schwarcz. "Lawyering in the Age of Artificial Intelligence." Minnesota Law Review, 109 (forthcoming 2024). Minnesota Legal Studies Research Paper No. 23-31. 

\item Choi, Jonathan H., and Daniel Schwarcz. "AI Assistance in Legal Analysis: An Empirical Study." Journal of Legal Education, 73 (forthcoming 2024). 

\item Dell'Acqua, Fabrizio, et al. "Navigating the Jagged Technological Frontier: Field Experimental Evidence of the Effects of AI on Knowledge Worker Productivity and Quality." Harvard Business School Technology \& Operations Mgt. Unit Working Paper 24-013, 2023.

\item Eloundou, Tyna, et al. "GPTs are GPTs: An Early Look at the Labor Market Impact Potential of Large Language Models." arXiv preprint arXiv:2303.10130, 2023.

\item Goldin, Claudia, and Lawrence F. Katz. "The Race between Education and Technology: The Evolution of US Educational Wage Differentials, 1890 to 2005." 2007.

\item Hackenburg, Kobi, et al. "Evidence of a Log Scaling Law for Political Persuasion with Large Language Models." arXiv preprint arXiv:2406.14508, 2024.

\item Kalchbrenner, Nal, and Phil Blunsom. "Recurrent Convolutional Neural Networks for Discourse Compositionality." arXiv preprint arXiv:1306.3584, 2013.

\item Kaplan, Jared, et al. "Scaling Laws for Neural Language Models." arXiv preprint arXiv:2001.08361, 2020.

\item Korinek, Anton and Donghyun Suh, “Scenarios for the Transition to AGI,” Technical
Report, National Bureau of Economic Research 2024

\item Sanchez Torron, M. Productivity in Post-Editing and in Neural Interactive Translation Prediction: A Study of English-to-Spanish Professional Translators. University of Auckland, 2017.

\item Jaime Sevilla et al. (2024), "Can AI Scaling Continue Through 2030?". 

\item Jaime Sevilla and Edu Roldán (2024), "Training Compute of Frontier AI Models Grows by 4-5x per Year". 

\item Jumper, J., Evans, R., Pritzel, A., et al. "Highly accurate protein structure prediction with AlphaFold." \textit{Nature}, vol. 596, no. 7873, 2021, pp. 583–589.

\item Shakked Noy, and Whitney Zhang. "Experimental Evidence on the Productivity Effects of Generative Artificial Intelligence." Science 381, 187-192 (2023). DOI:10.1126/science.adh2586.

\item Svanberg, Maja and Li, Wensu and Fleming, Martin and Goehring, Brian and Thompson, Neil, Beyond AI Exposure: Which Tasks are Cost-Effective to Automate with Computer Vision? (January 19, 2024). 

\item Toner-Rodgers, Aidan. "Artificial Intelligence, Scientific Discovery, and Product Innovation." (2024).

\item Ziegler, Albert, et al. "Productivity Assessment of Neural Code Completion." Proceedings of the 6th ACM SIGPLAN International Symposium on Machine Programming, 2022.

\end{itemize}

\FloatBarrier 

\pagebreak
\section{Appendix A}
\FloatBarrier 

In Appendix A, we report the regressions outlined in the Results section. The first result depicts the impact of participants having access to any AI model on the speed and quality of their completion. \\\

\begin{table}[htbp]
    \centering
    \caption{Impact on Time Taken and Quality of any AI Model Usage}
    \begin{tabular}{lcccccc}
        \toprule
        & \multicolumn{2}{c}{(1)} & \multicolumn{2}{c}{(2)} \\
        \midrule
        & \multicolumn{1}{c}{\textbf{Time}} & \multicolumn{1}{c}{\textbf{Grade}} & \multicolumn{1}{c}{\textbf{Time}} & \multicolumn{1}{c}{\textbf{Grade}} \\
        \midrule
        AI               & -186.8728   & 0.2038   & -188.6417   & 0.2438   \\
                         & (50.9583)   & (0.1408) & (50.5810)   & (0.1023) \\
        Arabic           &             &          & 65.3232    & -1.6788  \\
                         &             &          & (31.6758)  & (0.0641) \\
        Hindi            &             &          & -12.8134   & 0.5786   \\
                         &             &          & (32.0924)  & (0.0649) \\
        Task 2           &             &          & 137.2470   & -0.3586  \\
                         &             &          & (41.3007)  & (0.0836) \\
        Task 3           &             &          & 74.4368    & -0.4753  \\
                         &             &          & (41.3196)  & (0.0836) \\
        Task 4           &             &          & 177.9198   & -0.0784  \\
                         &             &          & (41.3100)  & (0.0836) \\
        Task 5           &             &          & 32.9588    & -0.3572  \\
                         &             &          & (41.3320)  & (0.0837) \\
        Constant         & 600.7017    & 4.5094   & 499.9389   & 5.1055   \\
                         & (49.0895)   & (0.1357) & (57.5884)  & (0.1165) \\
        \midrule
        Observations     & 1,500       & 1,499    & 1,500      & 1,499    \\
        R-squared        & 0.0089      & 0.0014   & 0.0297     & 0.4765   \\
        Adj R-squared    & 0.0082      & 0.0007   & 0.0251     & 0.4741   \\
        \bottomrule
    \end{tabular}
    \label{tab:regression}
\end{table}

\FloatBarrier 

The second set of regression results illustrate the impact of model training compute on the time taken by participants to complete tasks. \\\

{
\def\sym#1{\ifmmode^{#1}\else\(^{#1}\)\fi}
\begin{table}[!htbp]
\centering
\caption{Table 3: Scaling Laws for Time Taken}
\begin{tabular}{l*{4}{c}}
\hline\hline
                    &\multicolumn{1}{c}{(1)}&\multicolumn{1}{c}{(2)}&\multicolumn{1}{c}{(3)}&\multicolumn{1}{c}{(4)}\\
                    &\multicolumn{1}{c}{Time}&\multicolumn{1}{c}{Time}&\multicolumn{1}{c}{Time (log)}&\multicolumn{1}{c}{Time (log)}\\
\hline
Training Compute (log)   &      -51.35 &      -50.85 &     -0.0956  &     -0.0933  \\
                    &     (15.82)         &     (15.68)         &    (0.0375)         &    (0.0363)         \\
[1em]
Arabic              &                     &       67.49  &                     &     -0.0711         \\
                    &                     &     (31.91)         &                     &    (0.0739)         \\
[1em]
Hindi               &                     &      -28.95         &                     &      -0.564\\
                    &                     &     (32.38)         &                     &    (0.0750)         \\
[1em]
Task 2               &                     &       149.8&                     &       0.526\\
                    &                     &     (41.96)         &                     &    (0.0972)         \\
[1em]
Task 3               &                     &       84.69  &                     &       0.270 \\
                    &                     &     (41.78)         &                     &    (0.0967)         \\
[1em]
Task 4               &                     &       171.8&                     &       0.341\\
                    &                     &     (41.84)         &                     &    (0.0969)         \\
[1em]
Task 5               &                     &       47.25         &                     &       0.153         \\
                    &                     &     (41.70)         &                     &    (0.0966)         \\
\hline
Observations        &        1392         &        1392         &        1392         &        1392         \\
\hline\hline
\multicolumn{5}{l}{\footnotesize Standard errors in parentheses}\\

\end{tabular}
\end{table}
}

\FloatBarrier 

The third set of regression results depicts the impact of model training compute on the quality of tasks completed by participants. \\\

\begin{table}[htbp]
    \centering
    \caption{Scaling Laws for Quality}
    \begin{tabular}{lcc}
        \toprule
        & \multicolumn{1}{c}{(1)} & \multicolumn{1}{c}{(2)} \\
        \midrule
        & \multicolumn{1}{c}{\textbf{Grade}} & \multicolumn{1}{c}{\textbf{Grade}} \\
        \midrule
        logmodelcompute   & 0.2584      & 0.2510      \\
                          & (0.0442)    & (0.0322)    \\
        Arabic            &             & -1.6261     \\
                          &             & (0.0656)    \\
        Hindi             &             & 0.6024      \\
                          &             & (0.0665)    \\
        Task 2            &             & -0.3295     \\
                          &             & (0.0863)    \\
        Task 3            &             & -0.4503     \\
                          &             & (0.0859)    \\
        Task 4            &             & -0.1030     \\
                          &             & (0.0860)    \\
        Task 5            &             & -0.3730     \\
                          &             & (0.0857)    \\
        Constant          & -1.6922     & -0.9017     \\
                          & (1.0970)    & (0.8023)    \\
        \midrule
        Observations      & 1,391       & 1,391       \\
        R-squared         & 0.0240      & 0.4847      \\
        Adj R-squared     & 0.0233      & 0.4820      \\
        Root MSE          & 1.3865      & 1.0097      \\
        \bottomrule
    \end{tabular}
    \label{tab:regression3}
\end{table}

\FloatBarrier 

The fourth set of regression results illustrates the impact of model training compute on participant's total earnings per minute (inclusive of bonus payments for high-quality task completion). \\\

\begin{table}[htbp]
    \centering
    \caption{Scaling Laws for Total Earnings}
    \begin{tabular}{lcc}
        \toprule
        & \multicolumn{1}{c}{(1)} & \multicolumn{1}{c}{(2)} \\
        \midrule
        & \multicolumn{1}{c}{\textbf{Total epm}} & \multicolumn{1}{c}{\textbf{Total epm}} \\
        \midrule
        logmodelcompute   & 0.1921      & 0.1856      \\
                          & (0.0755)    & (0.0718)    \\
        Arabic            &             & -0.0258     \\
                          &             & (0.1462)    \\
        Hindi             &             & 1.4830      \\
                          &             & (0.1483)    \\
        Task 2            &             & -0.7958     \\
                          &             & (0.1922)    \\
        Task 3            &             & -0.5834     \\
                          &             & (0.1914)    \\
        Task 4            &             & -0.3019     \\
                          &             & (0.1917)    \\
        Task 5            &             & -0.3183     \\
                          &             & (0.1910)    \\
        Constant          & -3.5670     & -3.4693     \\
                          & (1.8725)    & (1.7888)    \\
        \midrule
        Observations      & 1,392       & 1,392       \\
        R-squared         & 0.0046      & 0.1035      \\
        Adj R-squared     & 0.0039      & 0.0990      \\
        Root MSE          & 2.3672      & 2.2514      \\
        \bottomrule
    \end{tabular}
    \label{tab:regression4}
\end{table}

\FloatBarrier 
\pagebreak
\section{Appendix B}

The following section offers the text description of all six tasks completed by participants in the experiment. \\\ 

\textbf{Task One:} \\\ 

\textit{Our calculations indicate that currently proposed U.S. policies to reduce pharmaceutical prices, though particularly beneficial for low-income and elderly populations, could dramatically reduce firms’ investment in highly welfare-improving R\&D. The U.S. subsidizes the worldwide pharmaceutical market. One reason is U.S. prices are higher than elsewhere. If each drug had a single international price across the highest-income OECD countries, and total pharmaceutical firm profits were held fixed, then U.S. prices would fall by half and every other country’s prices would increase (by 28 to 300\%). International prices would maintain firms’ R\&D incentives and more equitably share the costs of pharmaceutical research.} \\\ 

\textbf{Task Two:} \\\

\textit{One has only to look over the hedges of eastern England to agree with those who are predicting the worst harvest in living memory. What assessment has the Secretary of State made of the impact that will have on the wider rural economy—in particular, the availability and price of straw, which is vital for the livestock sector, and important commodities such as potatoes, which are likely to be under great pressure in terms of supply and price this autumn?} \\\ 

\textbf{Task Three:} \\\

\textit{At the time, college coaches handled their own endorsement deals separately from the universities they coached for, which often led to conflicts of interests between sponsors. When Jerry Claiborne became the Kentucky football head coach he signed a deal with Pepsi that included sideline rights in the stadium. Host attended a game that season with the largest Coca-Cola distributor in the state of Kentucky who noticed the Pepsi cups on the team's bench and grew confused.} \\\ 

\textbf{Task Four:} \\\ 

\textit{We provide evidence on the role of fairness for tax compliance: households are willing to pay more in taxes if they believe that other households are contributing their fair share. We conducted an information-disclosure natural field experiment in the context of property taxes in the United States. We induced exogenous shocks to households' perceptions about the average tax rate paid by other households. We find that a higher perceived average tax rate decreases the probability of filing a tax appeal. Translating our estimates into a money metric, we find that for each additional \$1 contributed by the average household, a taxpayer is willing to pay an extra \$0.43 in his or her own taxes.} \\\ 

\textbf{Task Five:} \\\ 

\textit{When I write my book of windows, I want to leave all the scenes open, so the wind can blow through. At the same time, I want to explain things. I want to explain those years when Mark and I were invisible to each other. When Mark disappeared into his relationship with Olivia, and I disappeared into my work–I was a fashion model, for a boutique agency in the city–and now we never talk about it, that period of our lives– one day we started referring to Mark’s relationship and my work in the past tense and that was that.}\\\ 

\textbf{Task Six:} \\\ 

\textit{This paper uses the responses to questions about charitable contributions from the Survey of Consumer Finances (SCF) between 1992 and 2022 to consider the rates of US households contributing money or time to charitable organizations. The fraction donating \$500 or more remained relatively constant over this period, with about 47\% answering they had donated in both 1991 and 2021. The fraction of households volunteering time declined consistently after 2005 from 34\% to 26\%.}

\end{document}